\documentclass[12pt]{article}
\usepackage{amsmath,amssymb,epsfig,youngtab}
\makeatletter \@addtoreset{equation}{section} \makeatother
\renewcommand{\theequation}{\thesection.\arabic{equation}}
\newcommand{\ba}{\begin{array}}
\newcommand{\ea}{\end{array}}
\newcommand{\beq}{\begin{equation}}
\newcommand{\eeq}{\end{equation}}
\newcommand{\bea}{\begin{eqnarray}}
\newcommand{\eea}{\end{eqnarray}}

\def\bce{\begin{center}}
\def\ece{\end{center}}

\def\nonu{\nonumber}

\def\pa{\partial}

\def\be{\beta}

\def\La{\Lambda}

\def\eps6{{\displaystyle \mathop{\epsilon}^{6}}{}}

\def\nab6{{\displaystyle \mathop{\nabla}^{6}}{}}

\def\0{{\sst{(0)}}}
\def\1{{\sst{(1)}}}
\def\2{{\sst{(2)}}}
\def\3{{\sst{(3)}}}
\def\4{{\sst{(4)}}}
\def\5{{\sst{(5)}}}
\def\6{{\sst{(6)}}}
\def\7{{\sst{(7)}}}
\def\8{{\sst{(8)}}}

\def\ba{\begin{array}}
\def\ea{\end{array}}
\def\beq{\begin{equation}}
\def\eeq{\end{equation}}
\def\be{\begin{equation}}
\def\ee{\end{equation}}

\def\Tr{\mathop{\rm Tr}}

\def\eps{\epsilon}

\def\ba{\begin{array}}
\def\ea{\end{array}}
\def\beq{\begin{equation}}
\def\eeq{\end{equation}}
\def\be{\begin{equation}}
\def\ee{\end{equation}}

\def\Tr{\mathop{\rm Tr}}

\def\eps{\epsilon}

\newcommand{\bean}{\begin{eqnarray*}}
\newcommand{\eean}{\end{eqnarray*}}

\begin{document}
\thispagestyle{empty} \addtocounter{page}{-1}
   \begin{flushright}
\end{flushright}

\vspace*{1.3cm}
 
 \centerline{ \Large \bf  Other Squashing Deformation and } 
\vspace{.3cm} 
\centerline{ \Large \bf  
${\cal N}=3$ Superconformal Chern-Simons Gauge Theory   } 
\vspace*{1.5cm}
\centerline{{\bf Changhyun Ahn }
} 
\vspace*{1.0cm} 
\centerline{\it  
Department of Physics, Kyungpook National University, Taegu
702-701, Korea} 
\vspace*{0.8cm} 
\centerline{\tt ahn@knu.ac.kr
} 
\vskip2cm

\centerline{\bf Abstract}
\vspace*{0.5cm}

We consider 
one of the well-known solutions in eleven-dimensional supergravity 
where the seven-dimensional Einstein space   
is given by a $SO(3)$-bundle over the
${\bf CP}^2$.
By reexaming the $AdS_4$ supergravity scalar potential, 
the holographic  
renormalization group flow 
from ${\cal N}=(0,1)$ $SU(3) \times SU(2)$-invariant UV fixed
point to ${\cal N}=(3,0)$ 
$SU(3) \times SU(2)$-invariant IR fixed point 
is reinterpreted. A dual operator in
three-dimensional  superconformal Chern-Simons matter theories
corresponding to this RG flow is described. 

\baselineskip=18pt
\newpage
\renewcommand{\theequation}
{\arabic{section}\mbox{.}\arabic{equation}}

\section{Introduction}

The ${\cal N}=6$ superconformal Chern-Simons matter theory
with gauge group $U(N) \times U(N)$ at level $k$ and with two
hypermultiplets in the bifundamental representations 
is found in \cite{ABJM}.
This gauge theory is described  
as the low energy limit of $N$ M2-branes probing 
${\bf C}^4/{\bf Z}_k$ singularity.
At large $N$-limit, this theory is dual to the eleven-dimensional M-theory on 
$AdS_4 \times {\bf S}^7/{\bf Z}_k$ where
the seven-sphere metric is realized 
as an ${\bf S}^1$-fibration over ${\bf
CP}^3$ \cite{NP}.
One of the main observations in \cite{ABJM} is to 
look at the special case of ${\cal N}=3$ superconformal Chern-Simons
matter theory with above particular gauge group, matter contents,
and particular choice of Chern-Simons levels of two gauge groups. Then
naive $SU(2)$ flavor symmetry appearing in the hypermultiplets
is enhanced to $SU(2)\times SU(2)$ symmetry that occurs in the whole
action of the theory and 
there exists a $SU(2)_R$ symmetry coming from the original ${\cal N}=3$
superconformal symmetry.  It turns out the full theory has ${\cal N}=6$
superconformal symmetry and the full scalar potential is invariant
under $SU(4)_R$ coming from this enhanced  ${\cal N}=6$
superconformal symmetry.

The simplest spontaneous compactification of the eleven-dimensional 
supergravity is the Freund-Rubin \cite{FR} compactification
to a product 
of $AdS_4$ spacetime and an arbitrary compact seven-dimensional Einstein 
manifold ${\bf X}^7$ 
of positive scalar curvature. 
The standard 
Einstein metric of the round seven-sphere
${\bf S}^7$ yields a vacuum with $SO(8)$ gauge 
symmetry and ${\cal N}=8$ supersymmetry. There exists a second 
squashed Einstein metric \cite{Jensen,BK} 
yielding a vacuum with $SO(5) \times 
SU(2)$ gauge symmetry and ${\cal N}=(1, 0)$ 
supersymmetry \cite{ADP,DNP}.
As suggested in \cite{GPPZ,FGPW},
in \cite{AR}, it was shown that the well-known spontaneous (super)symmetry 
breaking deformation from round ${\bf S}^7$ to squashed one is mapped to a 
renormalization group(RG) flow from ${\cal N}=(1, 0)$
$SO(5) \times SU(2)$-invariant fixed point in the UV 
to ${\cal N}=8$ $SO(8)$-invariant fixed 
point in the IR. In particular, 
the squashing 
deformation corresponds to an irrelevant operator at the UV
superconformal fixed point and a relevant operator at 
the IR (super)conformal fixed point respectively. Moreover the RG 
flow is described geometrically 
by a static domain wall 
which interpolates the two asymptotically $AdS_4$ spacetimes with round 
and squashed ${\bf S}^7$'s. For the different type of
compactifications where the internal space has nonzero four-form 
field strength, see also \cite{Ahn08,Ahn08-1}. 

One could ask \cite{Ahn0809} what happens when we perform 
${\bf Z}_k$-quotient \cite{ABJM} along the above whole RG flow \cite{AR}? 
Starting from the general, one parameter-family, 
metric for ${\bf CP}^3$ inside of seven-sphere and  
its seven-dimensional uplift metric on an ${\bf S}^1$-bundle over this
${\bf CP}^3$,  
the full eleven-dimensional metric with 
appropriate warp factors was constructed.
By analyzing the $AdS_4$ scalar potential, 
the holographic supersymmetric(or nonsupersymmetric) 
RG flow 
from ${\cal N}=(1, 0)$ 
$SO(5) \times U(1)$-invariant UV fixed
point to ${\cal N}=(6, 0)$ 
$SU(4)_R \times U(1)$-invariant IR fixed point was described in 
\cite{Ahn0809}. 
Each symmetry group is the subset of previous ones respectively.
That is, $SO(5) \times U(1)$ is contained in $SO(5) \times SU(2)$
and $SU(4)_R \times U(1)$ is contained in  $SO(8)$.
The squashing deformation corresponds to the singlet of ${\bf 20'}$
of $SU(4)_R$ and it is given by the quartic term for the matter 
fields transforming as fundamental representation under the $SU(4)_R$.
The dual Chern-Simons matter theory at the  ${\cal N}=(1, 0)$ 
$SO(5) \times U(1)$-invariant UV fixed
point is constructed in \cite{OP}. 

Now it is natural to ask that are there any other examples where 
some squashing deformation in ${\bf X}^7$
might provide a similar RG flow and one can think of 
some dual operator 
in three-dimensional boundary
conformal field theory? 
Yes, 
the manifold ${\bf X}^7 =  N^{0, 1, 0}_{\rm I}$ 
has been studied originally by Castellani and Romans \cite{CR} who 
identified this manifold as a particular coset manifold which
is specified by three integers. We consider the particular case where
$p=0, q=1$ and $r=0$ in $ N^{p, q, r}_{\rm I}$ manifold. 
There exists ${\cal N}=(3,0)$  
supersymmetry with $SU(3)\times SU(2)$ gauge symmetry 
or ${\cal N}=(1,0)$ supersymmetry with $SU(3)\times U(1)$ gauge 
symmetry(See also \cite{Castellani}). 
Moreover, Page and Pope \cite{PP} 
have completed the coset manifold construction 
by showing existence of another family of Einstein manifold, 
$N^{0,1,0}_{\rm II}$, which can be obtained from geometric squashing 
of the $N^{0,1,0}_{\rm I}$, retains the same gauge group 
$SU(3) \times SU(2)$ but instead preserves 
${\cal N} = (0,1)$ supersymmetry.
As in the case of seven-sphere ${\bf S}^7$, 
the scalar field corresponding to the squashing deformation acquires a 
nonzero vacuum expectation value leading to (super)-Higgs mechanism. 
With left-orientation the squashing interpolates between a 
${\cal N}=3$ supersymmetric vacuum and another with ${\cal N}=0$ supersymmetry. 
With right-orientation it interpolates between a nonsupersymmetric
vacuum and a supersymmetry restored one with ${\cal N}=1$
supersymmetry \cite{AR1}.

On the other hand, 
in \cite{BFFMZ}, the corresponding ${\cal N}=3$ dual gauge theory  
has gauge group $SU(N) \times SU(N)$ with ``three'' hypermultiplets
transforming as a ``triplet'' under the $SU(3)$ flavor symmetry
which is nothing but one of the global symmetries for 
$N_{\rm I}^{0,1,0}$ manifold \footnote{We emphasize that this theory
is based on Yang-Mills plus Chern-Simons theory with chiral multiplets, contrary to
\cite{ABJM} in which the theory is described as Chern-Simons 
theory with chiral multiplets. For example, the kinetic terms in \cite{BFFMZ}
contain those for the vector multiplet as well as the Chern-Simons
term and those for the chiral multiplets. However, in \cite{ABJM}, there is no
kinetic term for any of the fields in the vector multiplet.   }. 
In terms of ${\cal N}=2$ superfields, these hypermultiplets 
can be reorganized as two sets of chiral superfields.
For the color representation, one of these transforms as 
$({\bf N, \overline{N}})$ and the other transforms as
$({\bf \overline{N}, N})$. 
Furthermore, these two superfields transform as a doublet of $SU(2)_R$
which is the $SU(2)$ factor in the remaining global symmetry of
$N_{\rm I}^{0,1,0}$ manifold. As we mentioned before, 
$SU(2)_R$ corresponds to the ${\cal N}=3$ superconformal symmetry.
After integrating out two adjoint fields of the theory, 
the effective quartic 
superpotential can be obtained. The coefficient is determined by the 
${\cal N}=3$ supersymmetry but breaks ${\cal N}=4$ supersymmetry.  For
the general discussion on ${\cal N}=3$ superconformal Chern-Simons
matter theory, see \cite{GY}.
The complete ${\cal N}=3$ Kaluza-Klein spectrum is found in 
\cite{Termonia} and its $OSp(3|4)$ multiplet structure is further explained 
in \cite{FGT}. 

Later, Billo, Fabbri, Fre, Merlatti and Zaffaroni
\cite{BFFMZ1}(See also \cite{FF}) have
constructed an ${\cal N}=3$ long 
massive spin 3/2 multiplet with conformal
dimension 3 from the massless ${\cal N}=3$ graviton multiplet. These
two are connected to ``shadow'' relation: fields of different type,
spin and mass are linked by a relation which determines the mass of
the one as a function of the other. 

In this paper,
we will be studying the known example of Kaluza-Klein 
supergravity vacua and reinterpret it in terms of three-dimensional
(super)conformal field theories and associated RG flows. 
We will be exploring the Freund-Rubin type spontaneous
compactification on $AdS_4 \times {\bf X}^7$. 
For M2-branes on an eight-dimensional
manifold, the near-horizon geometry ${\bf X}^7$ is expected to change as the 
M2-branes are placed at or away a conical singularity of the manifold 
\cite{AFHS, MP}. More specifically, we will consider ${\bf X}^7$ being
3-Sasaki  holonomy manifold,
describing near-horizon geometry of M2-branes at relevant conical
singularities.

When the work of \cite{AR1} was completed at that time, 
it was not possible to analyze the gauge theory description because
the Kaluza-Klein spectrum of eleven-dimensional supergravity was not complete.
Later, in \cite{BFFMZ1}, they have found more mass spectrum in the
eleven-dimensional supergravity side that includes the harmonics of
the Lichnerowicz scalar with conformal dimension 4.
Then one can identify the corresponding 
fluctuation spectrum for the scalar fields around 
${\cal N}=3$ fixed point. 

The aim of this paper is to 1) recapitulate the effective scalar
potential described in \cite{AR1} with only breathing mode and
squashing mode, and 2) analyze more both the mass spectrum in the
eleven-dimensional supergravity and the corresponding Chern-Simons 
gauge theory
operator which gives rise to the squashing deformation, by analyzing 
the results of \cite{BFFMZ1}.

In section 2,
we describe the seven-dimensional Einstein space($N_{\rm I}^{0,1,0}$) and 
its squashed version($N_{\rm II}^{0,1,0}$)  
compactification vacua in
eleven-dimensional supergravity. 
The effective four-dimensional scalar potential looks similar to 
the one for seven-sphere and the two critical points have nonzero scalar
fields. However, the ratio of the squashing parameter at these two
critical values(which is equal to 
$1/5$) is the same as the one in seven-sphere case. 

In section 3,
the squashing deformation of  each vacua
is described by an irrelevant operator at the ${\cal N}=(3, 0)$ 
conformal
fixed point and a relevant operator at the ${\cal N}=(0, 1)$
conformal fixed points. The RG flow is described in $AdS_4$
supergravity by a static domain wall interpolating between 
these two vacua. We identify the corresponding operator 
in the boundary conformal field theory in three dimensions by looking
at the observations of \cite{BFFMZ1} \footnote{Recently, the 
${\cal N}=3$ superconformal Chern-Simons quiver theories are
constructed in \cite{JT} but these theories do not contain the
seven-dimensional Einstein manifold we are considering here. See also
other relevant paper \cite{LY}. For the earlier studies on 
${\cal N}=3$ superconformal Chern-Simons theories, there are also 
some works in \cite{ZK,Kao,KLL}.}.

\section{Two seven-dimensional Einstein spaces }

A generic eleven-dimensional metric 
interpolating between two seven-dimensional Einstein spaces
with an arbitrary four-dimensional spacetime metric maybe 
written as 
\bea
\frac{\overline{ds}^2}{R^2} & = &   
e^{- 7u} g_{\alpha \beta} d x^\alpha d x^\beta
 +       e^{2u+3v}  \left[ d \mu^2 + 
 \frac{1}{4} \sin^2 \mu 
( \sigma_1^2 + \sigma_2^2 + \cos^2 \mu \; \sigma_3^2 )   \right] 
 \nonu \\
& + &     e^{2u -4v} \left[ 
\left( \Sigma_1 -\cos \mu \; \sigma_1 \right)^2 
 + 
 \left(  \Sigma_2 -\cos \mu \; \sigma_2  \right)^2 
+ \left(  \Sigma_3 -\frac{1}{2} (1+ \cos^2 \mu) \sigma_3  \right)^2 
\right], 
\label{11dmetric}
\eea
where the three real left-invariant one-forms satisfy 
the $SU(2)$ algebra $d \sigma_i = -\frac{1}{2} \epsilon_{ijk} 
\sigma_j \wedge \sigma_k$ and 
those on the manifold $SO(3)$ which is necessary for the regularity of
the metric have 
$d \Sigma_i = -\frac{1}{2} \epsilon_{ijk} 
\Sigma_j \wedge \Sigma_k$ \footnote{ 
They are given by 
$
\sigma_1   =  \cos \psi d \theta + \sin \psi \sin \theta d \phi$,  
$ \sigma_2   =  -\sin \psi d \theta + \cos \psi \sin \theta d \phi$
 and $ \sigma_3   =  d \psi + \cos \theta d \phi$
and
similarly $SO(3)$ one-forms are
given by $
\Sigma_1   =  \cos \gamma d \alpha + \sin \gamma \sin \alpha d \beta$,  
$ \Sigma_2   =  -\sin \gamma d \alpha + \cos \gamma \sin \alpha d \beta$, 
and $
\Sigma_3   =  d \gamma + \cos \alpha d \beta$ \cite{DNP86,Ahn02mpla}.}.
The coordinate $\mu$ and $SU(2)$ one-forms $\sigma_i$ are the same as 
${\bf CP}^2$ metric and corresponding isometry is characterized by 
$SU(3)$. Then the seven-dimensional 
$N^{0,1,0}$ space is a nontrivial $SO(3)$ bundle over
${\bf CP}^2$ and the isometry group for this space 
is $SU(3) \times SU(2)$ \cite{PP}. 
The scalar fields $u(x)$ for the breathing mode 
and $v(x)$ for squashing mode depend on the four-dimensional spacetime.
Moreover, the squashing is parametrized by \cite{Page}
\bea
\lambda^2 \equiv e^{- 7v}.
\label{squ}
\eea
The parameter $R$ measures the overall radius of curvature. 
The gauge fields are given by $A_1= \cos \mu \; \sigma_1, 
A_2 = \cos \mu \; \sigma_2$ and $  
A_3 = \frac{1}{2} (1+ \cos^2 \mu) \sigma_3$. 

Spontaneous compactification of M-theory to 
$AdS_4 \times {\bf X}^7$ is obtained from near-horizon
geometry of $N$ coincident M2-branes and the nonvanishing flux of 
four-form field strength of the Freund-Rubin \cite{FR} is given by
\bea
\overline{F}_{\alpha \beta \gamma \delta}
= Q e^{-7 u } \overline{\epsilon}_{\alpha \beta \gamma \delta}
= Q e^{-21 u} \epsilon_{\alpha \beta \gamma \delta}. 
\label{fieldstrength}
\eea
Here the Page charge \cite{Page,DNP86}
defined by
$
Q \equiv \pi^{-4} 
\int_{{\bf X}^7} ({}^*\overline{F} + \overline{C} \wedge \overline{F})
$ is related to the total number of M2-branes through 
$Q = 96 \pi^2 N \ell_p^6$.
Then the eleven-dimensional Einstein equation with
(\ref{fieldstrength}) 
provides the following 
Ricci tensor components \cite{Page,Ahn0809}
\bea
\overline{R}^{\alpha}_{\beta} = -\frac{4}{3} Q^2 e^{-14 u} 
\delta^{\alpha}_{\beta}, \qquad 
\overline{R}^{a}_{b} = \frac{2}{3} Q^2 e^{-14 u} 
\delta^{a}_{b},  \qquad
\overline{R}^{\alpha}_{b} = \overline{R}^{a}_{\beta} = 0.
\label{result1}
\eea

On the other hand,  the Ricci tensor components can be obtained from
the following orthonormal basis which can be read off from the
eleven-dimensional metric (\ref{11dmetric})
\bea
e^1 & = &   e^{- \frac{7}{2} u} \sqrt{g_{11}(x)} dx^1, 
\qquad e^2   =    e^{- \frac{7}{2} u} \sqrt{g_{22}(x)} dx^2, \nonu \\
e^3 & =  &  e^{- \frac{7}{2} u} \sqrt{g_{33}(x)} dx^3, \qquad
e^4    =     e^{- \frac{7}{2} u} \sqrt{g_{44}(x)} dx^4, \nonu \\
e^5  & = &    e^{u+\frac{3}{2} v}  d \mu, \qquad
e^6  =   \frac{1}{2} e^{u+\frac{3}{2} v} 
\sin \mu \; \sigma_1, \nonu \\
e^7  & = &   \frac{1}{2} e^{u+\frac{3}{2} v} \sin \mu
\; \sigma_2, \qquad
e^8  =   \frac{1}{2} e^{u+\frac{3}{2} v} \sin \mu \cos \mu 
\; \sigma_3, \nonu \\  
e^9 & = &  e^{u -2v} 
\left( \Sigma_1 - \cos \mu \; \sigma_1 \right), \qquad
e^{10}  =   e^{u-2v} \sin \mu 
\left(  \Sigma_2 - \cos \mu \; \sigma_2
\right), 
\nonu \\
e^{11} & = &  e^{u-2v} 
\left[  \Sigma_3 -\frac{1}{2} (1+ \cos^2 \mu) \sigma_3  \right]. 
\label{frame}
\eea
The results for Ricci tensor 
in the basis of (\ref{frame}) are summarized by 
\bea
\overline{R}^{\alpha}_{\beta}  & = &  e^{7u} \left( R^\alpha_\beta +
\frac{7}{2} \delta_{\beta}^{\alpha} u^{;\gamma}_{;\gamma}
-\frac{63}{2}
u^{;\alpha} u_{;\beta} -21 v^{;\alpha} v_{;\beta} \right),
\nonu \\
\overline{R}^{5}_{5} & = & 6 e^{-2u-3v} - 6 e^{-2u-10v} -
e^{7u} \left( u^{;\alpha}_{;\alpha} +
\frac{3}{2} v^{;\alpha}_{;\alpha} \right)
=\overline{R}^{6}_{6} =\overline{R}^{7}_{7}=\overline{R}^{8}_{8},
\nonu \\
\overline{R}^{9}_{9} & = & \frac{1}{2} e^{-2u+4v} + 4 e^{-2u-10v} -
e^{7u} \left( u^{;\alpha}_{;\alpha} -2 v^{;\alpha}_{;\alpha} \right)
=\overline{R}^{10}_{10} =\overline{R}^{11}_{11}.
\label{result}
\eea
Note that the only first terms in $\overline{R}^{5}_{5}$
and $\overline{R}^{9}_{9}$ are different from the one for squashed
${\bf S}^7$ space \cite{Page,Ahn0809}
\footnote{
When the scalar fields $u(x)$ and $v(x)$ are constant, then 
$
\overline{R}^{5}_{5}  =  6 - 6 e^{-7v}=6 - 
6 \lambda^2 =\overline{R}^{6}_{6} =\overline{R}^{7}_{7}=
\overline{R}^{8}_{8}$ and 
$ \overline{R}^{9}_{9}  =  4 e^{-7v} +\frac{1}{2} e^{7v} = 4 \lambda^2 +
\frac{1}{2\lambda^2} = \overline{R}^{10}_{10}
=\overline{R}^{11}_{11}$ where (\ref{squ}) is used.
Then $\lambda^2 =\frac{1}{2}$ corresponds to the Einstein metric 
in \cite{CR} and $\lambda^2=\frac{1}{10}$ corresponds to 
the squashed Einstein metric \cite{PP}.    }.

Substituting the last two relations in (\ref{result}) into 
(\ref{result1}) implies 
the field equations  for the breathing mode $u(x)$ and the squashing
mode $v(x)$ as follows:
\bea
 u^{;\alpha}_{;\alpha}  & = & 
\frac{3}{14} e^{-9u+4v} +\frac{24}{7} e^{-9u-3v} -
\frac{12}{7} e^{-9u-10v}-
\frac{2}{3} Q^2 e^{-21u}, \nonu \\
v^{;\alpha}_{;\alpha}  & = & - e^{-9u+4v} +12 
e^{-9u-3v} -20 e^{-9u-10v}.
\label{result2}
\eea
Note that the only first two terms in $ u^{;\alpha}_{;\alpha}$
and $v^{;\alpha}_{;\alpha}$ are different from the one for squashed
${\bf S}^7$ space \cite{Page,Ahn0809}.
The vanishing of the right hand side of second equation of (\ref{result2}) implies
that either $v = v_1 = \frac{1}{7} \ln 2$ or 
$v=v_2 = {1 \over 7} \ln 10$.  
Furthermore, substituting the field equation for $u(x)$ in 
(\ref{result2}) into the first equation of (\ref{result})
together with (\ref{result1}) implies  the following four-dimensional
Ricci tensor 
\bea
R^{\alpha}_{\beta}  =  {63 \over 2} u^{;\alpha} u_{;\beta}
+ 21 v^{;\alpha} v_{;\beta} 
+ \delta^{\alpha}_{\beta} e^{-9u}
\left( - \frac{3}{4} e^{4v} - 12 e^{-3v} +6 e^{- 10v} + 
Q^2 e^{-12u} \right).
\label{result3}
\eea

Now the field equations (\ref{result2}) and (\ref{result3})
are equivalent to the Euler-Lagrange equations for the following 
effective Lagrangian 
\bea
{\cal L}= \sqrt{-g} \left[ R - \frac{63}{2} ( \pa u )^2 -
21 ( \pa v )^2 -V(u, v) \right]
\label{Lag}
\eea
with the scalar potential \footnote{This form for the scalar potential was
observed also previously in \cite{CFLMN}.}
\bea
V(u, v)= e^{-9u} \left( - \frac{3}{2} e^{4v} -24 e^{-3v}+12 e^{-10v} 
+  2  Q^2 e^{-12u} \right).
\label{potential}
\eea
Note that the first two terms in (\ref{potential}) are 
different from the one for squashed
${\bf S}^7$ space \cite{Page,AR,Ahn0809}.

One analyzes two vacua of this scalar potential as follows: 
\bea
N_{\rm I}^{0,1,0} \,\, \quad : \quad u = u_1 &=& {1 \over 12} \ln
\left( \frac{2^{\frac{10}{7}}}{3^2}  Q^2 \right),
\qquad v = v_1 = \frac{1}{7} \ln 2, \qquad \qquad(\lambda^2 = \frac{1}{2}),
\nonu \\
\Lambda_1 &=& -9 \left\vert { 2 Q^3 \over 3} \right\vert^{-\frac{1}{2}},
\nonu
\eea
and 
\bea
N_{\rm II}^{0,1,0} \quad : \quad 
u = u_2 &=& {1 \over 12} \ln \left(\frac{10^{\frac{10}{7}}}{3^4}  Q^2 \right), \qquad
v = v_2 = {1 \over 7} \ln 10, \qquad \left(\lambda^2 = {1 \over 10} \right), 
\nonu \\
\Lambda_2 &=& - 3^{6}\cdot 5^{-2}
\left\vert { 10 Q^3 } \right\vert^{-\frac{1}{2}}.
\nonu
\eea
The two supergravity solutions are classically stable under the
changes of the size and squashing parameter of seven-dimensional space 
\cite{Yasuda84}.
The $N_{\rm I}^{0,1,0}$ is a saddle point,  corresponds to a
minimum along the $v$-direction and is invariant under the $SU(3)
\times SU(2)$ isometry group while $N_{\rm II}^{0,1,0}$
is a maximum and is invariant under the same $SU(3) \times SU(2)$ isometry group.
The left-handed seven-dimensional space $N_{\rm I,L}^{0,1,0}$ 
gives rise to a theory 
with  ${\cal N}=3$ supersymmetry  
while 
the right-handed seven-dimensional space $N_{\rm I,R}^{0,1,0}$ 
gives rise to a theory 
with  no supersymmetry(${\cal N}=0$) \cite{PP}.

Moreover, 
the left-handed squashed seven-dimensional space $N_{\rm II,L}^{0,1,0}$ 
gives rise to a theory 
with  no supersymmetry  
while 
the right-handed squashed seven-dimensional space $N_{\rm II,R}^{0,1,0}$ 
gives rise to a theory 
with  ${\cal N}=1$ supersymmetry.
That is, with the choice of left-handed orientation of 
$N_{\rm L}^{0,1,0}$, one regards the $\lambda^2 =\frac{1}{2}$ metric 
as giving the unbroken vacuum state with ${\cal N}=3$ supersymmetry 
which can be broken spontaneously to $\lambda^2=\frac{1}{10}$ metric
yielding ${\cal N}=0$ supersymmetry. 
On the other hand, with the choice of opposite orientation of
${N_{\rm R}^{0,1,0}}$ the $\lambda^2=\frac{1}{10}$ metric provides a vacuum
state with ${\cal N}=1$ supersymmetry which can be broken to 
the $\lambda^2 =\frac{1}{2}$ metric with ${\cal N}=0$ supersymmetry.

Therefore, for the squashing with left-handed orientation, the RG flow
interpolates between the boundary conformal field theories with ${\cal
N}=3$ and ${\cal N}=0$ supersymmetry while for the squashing with 
right-handed orientation, the RG flow interpolates between conformal 
field theories with ${\cal N}=0$ and ${\cal N}=1$ \footnote{It is
known that the $N_{\rm I}^{0,1,0}$ can be written as an
$U(1)$-fibration over a Kahler-Einstein six-manifold that is an ${\bf
S}^2$-bundle over ${\bf CP}^2$ \cite{GLMW}. For the general squashed case 
$N^{0,1,0}$ including $N_{\rm II}^{0,1,0}$, 
one can also get $U(1)$-fibration over a six-manifold by reorganizing
the last line of (\ref{11dmetric}):
\bea
&& ds_7^2  = 
2 \lambda^2 \left[ -\frac{1}{2} d \psi -\frac{1}{2} \cos \mu \sin
  \theta (\cos \phi \; \sigma_1 -\sin \phi \; \sigma_2) 
 +\frac{1}{4} \cos
  \theta(
1+ \cos^2 \mu) \sigma_3 -\frac{1}{2} \cos \theta d \phi \right]^2 
 \nonu \\
&& +  \frac{\lambda^2}{2} \left(
 \left[ d \theta -\cos \mu ( \sin \phi  \sigma_1 + \cos \phi  \sigma_2  ) 
\right]^2  
 +  
\sin^2 \theta \left[ d \phi -\cos \mu \cot \theta( \cos
  \phi \sigma_1 -\sin \phi \sigma_2) -\frac{1}{2} 
(1+ \cos^2 \mu) \sigma_3  \right]^2  \right) \nonu \\
&& +   \frac{1}{2} \left[
d \mu^2 + 
 \frac{1}{4} \sin^2 \mu 
( \sigma_1^2 + \sigma_2^2 + \cos^2 \mu \; \sigma_3^2 ) \right],
\nonu
\eea
where we choose the same parametrization given in \cite{GLMW}. The
Euler angles $\theta, \phi$ and $\psi$ correspond to the one-forms 
$\Sigma_i$. Of course, for the case of $\lambda^2=\frac{1}{2}$, this
metric leads to the one in \cite{GLMW}.  }.

\section{Three-dimensional (super)conformal field theories }

Using the previous results  on the Kaluza-Klein spectrum
under squashing deformations, 
an operator giving rise to a RG flow
associated with the supersymmetry breaking will be identified and 
it turns out that the operator is relevant at the 
$N_{\rm II}^{0,1,0}$ fixed point and irrelevant at the
$N_{\rm I}^{0,1,0}$ fixed point with same $SU(3) \times SU(2)$
symmetry groups.

$\bullet$ $SU(3) \times SU(2)$-invariant conformal fixed point

Let us consider the harmonic 
fluctuations of spacetime metric and $u(x)$ and $v(x)$ scalar fields
around $AdS_4 \times N_{\rm I}^{0,1,0}$.
Following \cite{Page,AR,Ahn0809}, it is more convenient 
to rewrite (\ref{Lag}) in terms of the unrescaled 
M-theory metric $\overline{g_{\alpha \beta}} = 
e^{-7u} g_{\alpha \beta}$  in (\ref{11dmetric}):
\bea
{\cal L} 
 =   \sqrt{-\overline{g}} e^{7u} \left[ \, \overline{R} -
2\overline{\La_1} - 
105 ( \pa u )^2 - 21 ( \pa v )^2 -2 \overline{V_1}(u, v) \, \right],   
\label{Lag1}
\eea 
where the scalar potential is written as
\bea
\overline{V_1}(u, v)= 
-\overline{\La_1} \left[ 1-\frac{1}{4} e^{-2(u-u_1)} ( e^{4(v-v_1)}+8e^{-3(v-v_1)}-
2e^{-10(v-v_1)} ) +\frac{3}{4} e^{-14(u-u_1)} \right]
\nonu
\eea
in which the un-rescaled cosmological constant 
$\overline{\La_1}
= e^{7u_1} \La_1 =  \frac{1}{2} e^{7u_1} V(u_1,v_1)$ is given by 
\bea
\overline{\La_1} \equiv - 3 m_1^2 {1 \over \ell_p^2}= 
-3 \left( \frac{| Q|}{6} \right)^{-\frac{1}{3}} {1 \over \ell_p^2}
\qquad {\rm where} \qquad m_1 = {1 \over \overline{r_{\rm IR}}}.
\label{lambda}
\eea
Here
$\overline{r_{\rm IR}}$ is related to $N$ 
and Planck scale $\ell_{\rm p}$ as 
$\overline{r_{\rm IR}} = \ell_{\rm p} {1 \over 2} ( 32 \pi^2  N)^{1/6}$.

By rescaling the scalar fields from the kinetic terms in the
Lagrangian (\ref{Lag1}) as 
\bea
\sqrt{210} u \equiv \overline{u}, \qquad
\sqrt{42} v \equiv \overline{v},
\nonu
\eea 
one obtains the
fluctuation spectrum for $\overline{v}$-field 
around the $N_{\rm I}^{0,1,0}$ 
which takes a positive value:
\bea
M_{\overline{v} \overline{v}}^2 (N_{\rm I}^{0,1,0}) 
= \left( \frac{\pa^2} {\pa \overline{v}^2} 2 \overline{V_1}
\right)_{\overline{u}=\overline{u_1}, \overline{v}=\overline{v_1}} = 
-
\frac{4}{3} \overline{\La_1} \ell_p^2=  4 m_1^2
\label{massmass}
\eea
where the relation (\ref{lambda}) is used.

Recall that in the compactification of $AdS_4 \times {\bf S}^7$,  
the $\overline{v}$-field represents the squashing of ${\bf
S}^7$ and hence ought to correspond to ${\bf 300}$(that is the
Young tableaux $\tiny\yng(2,2)$ of $SO(8)$ or  the  $SO(8)$
Dynkin label is given by $( {\bf 0, 2, 0, 0})$): the lowest mode of the 
transverse, traceless symmetric tensor representation.
The branching rule of the representation ${\bf 300}$ in
terms of 
$SO(7)$ Dynkin labels is given
by \cite{Slansky,Petera}  
$
{\bf 300}(\tiny\yng(2,2))  \rightarrow   
{\bf 27}(\tiny\yng(2)) \oplus {\bf 105}(\tiny\yng(2,1)) 
\oplus {\bf 168}(\tiny\yng(2,2))
$
and the Lichnerowicz operator $\Delta_L$ in $SO(7)$ representation
becomes ${\bf 27}$ \cite{DNP86} 
which has a Dynkin label $({\bf 2, 0, 0})$ and its
branching rule in
terms of maximal subgroup $SU(4)$ Dynkin labels is given
by 
$
{\bf 27}(\tiny\yng(2))  \rightarrow   
{\bf 1}(\tiny\yng(1,1,1,1)) \oplus {\bf 6}(\tiny\yng(2,1,1)) 
\oplus {\bf 20'}(\tiny\yng(2,2))$.
Note that the ${\bf 20'}$ in $SU(4)$ is represented by a 
traceless symmetric matrix and  
the squashing  should
correspond to nonzero expectation value for the ${\bf 20'}$ of $SU(4)$.
Then the branching rule of this ${\bf 20'}$  in
terms of $SU(3)$ which is a global symmetry for $N_{\rm I}^{0,1,0}$ 
is given by 
\bea
{\bf 20'}(\tiny\yng(2,2))  \rightarrow   
{\bf 6}(\tiny\yng(2)) \oplus \overline{{\bf 6}}(\tiny\yng(2,2)) 
\oplus {\bf 8}(\tiny\yng(2,1)).
\nonu
\eea
Moreover, 
 the $ \overline{{\bf 6}}$ in $SU(3)$ is represented by a 
traceless symmetric matrix and  
the squashing should eventually
correspond to nonzero expectation value for the $ \overline{{\bf 6}}$
of $SU(3)$.

According to the nice observations of 
\cite{BFFMZ1}, the harmonics of Lichnerowicz scalars with conformal
dimension $\Delta =4$ has been obtained. 
The harmonics are eigenfunctions of the 
Lichnerowicz operator with eigenvalue $M_{{\bf 200}}=96 m^2$ in
their notation.
Recall that the representation $(\bf 2, 0, 0)$ refers to 
$SO(7)$ representation $\bf 27$. 
With the help of \cite{DF}, the value of $M_{{\bf 200}}$ can be
obtained from $M_{{\bf \frac{3}{2} \frac{1}{2} \frac{1}{2}}}$ which
is an eigenvalue for Rarita Schwinger operator and it turns out
$-16$.  The representation $(\bf \frac{3}{2}, \frac{1}{2},
\frac{1}{2})$ 
refers to $SO(7)$ representation $\bf 48$. 
The explicit form is given by 
$M_{{\bf 200}} = (M_{{\bf \frac{3}{2} \frac{1}{2} \frac{1}{2}}} +4)(
M_{{\bf \frac{3}{2} \frac{1}{2} \frac{1}{2}}} +8) m^2$.
By plugging $M_{{\bf \frac{3}{2} \frac{1}{2} \frac{1}{2}}}=-16$,
one gets  $M_{{\bf 200}}=96 m^2$.
Note that $m^2$ is mass-squared parameter of a given $AdS_4$ spacetime
and the mass  of a scalar field $\phi$ in \cite{DF} is defined 
as $(\Delta_{AdS} + M_{{\bf 200}} -32m^2)\phi=0$.
Then $96m^2$ goes to $64m^2$ by subtracting $32m^2$ 
and by dividing out 16 further in order to compare with the
usual normalization in AdS/CFT correspondence,
one arrives that 
the correct result for the mass-squared is $4m^2$ which is the same
normalization used in \cite{AR1}. For example, 
see also the recent paper \cite{KKM} for the normalization between the
conformal dimension and the mass-squared term.
Therefore, the mass-squared for the representation $\overline{\bf 6}$
of $SU(3)$ is given by
\bea 
M_{\overline{\bf 6}}^2 = 4 m_1^2,
\nonu
\eea
which is exactly equal to (\ref{massmass}). Note that 
the assignment of $SU(2)_R$ isospin $J$ is related to the $SU(3)$
assignment and the R-symmetry group is the maximal $SU(2)$ subgroup of 
$SU(3)$. Under this embedding, the $SU(3)$ representations decompose
into as follows: 
\bea
{\bf 1, \overline{1}} \rightarrow J=0, \qquad
{\bf 3, \overline{3}} \rightarrow J=1, \qquad
{\bf 6, \overline{6}} 
\rightarrow J=0 \oplus J=2.
\nonu
\eea 
Or we have
$\overline{{\bf 6}}(\tiny\yng(2,2))  = 
\overline{{\bf 1}}(\tiny\yng(1,1)) 
\oplus \overline{{\bf 5}}(\tiny\yng(2))$ under the breaking of $SU(3)$
into the $SU(2)_R$.

One concludes that, in three-dimensional conformal field 
theory with ${\cal N} = 3$ supersymmetry, the $SU(3) \times SU(2)$ symmetric 
left-handed squashing should be an irrelevant perturbation 
of conformal dimension 
$\Delta = 4$. Note that this gives a 
nonsupersymmetric  theory for the 
right-handed orientation for seven-dimensional Einstein manifold
$N_{\rm I, R}^{0,1,0}$.

$\bullet$ $SU(3) \times SU(2)$-invariant conformal fixed point

Due to the skew-whipping, the theory will be either 
left-squashed  $N_{\rm II,L}^{0,1,0}$ with ${\cal
  N}=0$ supersymmetry or 
right-squashed  $N_{\rm II,R}^{0,1,0}$ with ${\cal
  N}=1$ supersymmetry.
The isometry of the squashed seven-dimensional Einstein manifold 
is given by $SU(3) \times
SU(2)$.
In terms of the unrescaled M-theory metric, the Lagrangian 
(\ref{Lag}) can be rewritten as
\bea
{\cal L}  = 
\sqrt{-\overline{g}} e^{7u} \left[ \, \overline{R} -
2\overline{\La_2}  - 
105 ( \pa u )^2 -
21 ( \pa v )^2 -2 \overline{V_2}(u, v) \, \right],   
\nonu
\eea 
where the scalar potential is given by
\bea
\overline{V_2}(u, v)  = 
 -\overline{\La_2} \left[ 1-\frac{1}{36} e^{-2(u-u_2)} 
\left( 25 e^{4(v-v_2)}+ 40 e^{-3(v-v_2)}- 2 e^{-10(v-v_2)} \right)  
+\frac{3}{4} e^{-14(u-u_2)} \right],
\nonu
\eea
and the un-rescaled cosmological constant $\overline{\La_2} =
e^{7 u_2} \La_2 =\frac{1}{2} e^{7u_2} V(u_2,v_2)$ is given by
\bea
\overline{\La_2}  &\equiv& - 3 m_2^2 {1 \over \ell_{\rm p}^2}
= - 3\left[ 3^{\frac{7}{3}} \cdot 5^{-\frac{2}{3}}  
\left( {\vert  Q \vert  \over 6} 
\right)^{-\frac{1}{3}} \right]
{1 \over \ell_{\rm p}^2}, \quad {\rm where}
\quad m_2 = {1 \over \overline{ r_{\rm UV}}   }.
\nonu
\eea

The mass spectrum of the $\overline{v}(x)$ field is calculated 
similarly
\bea
M^2_{\overline{v} \overline{v}}(N_{\rm II}^{0,1,0}) 
\equiv \left( \frac{\pa^2}
{\pa \overline{v}^2} 2 \overline{V_2} \right)_{\overline{u}=
\overline{u_2}, \overline{v}=\overline{v_2}} \,\, = \,\,  
\frac{20}{27} \overline{\La_2} \ell^2_{\rm p} = -\frac{20}{9} m_2^2.
\label{mass2}
\eea
From the 
mass formula for the $SU(3) \times SU(2)$ 
representation
and the
eigenvalues of the Lichnerowicz operator, 
one should obtain the mass-squared for the singlet  
as follows:
$
M_{(\overline{\bf 1}, \overline{\bf 1})}^2 =  -\frac{20}{9} m_2^2$,
and this coincides with (\ref{mass2}).
The perturbation that corresponds to squashing
around $N_{\rm II}^{0,1,0}$ has a scaling dimension either 
$\Delta = 4/3$ or $5/3$ and hence corresponds to a relevant operator.

We gave a nonzero expectation value to a supergravity scalar in the 
$\overline{\bf 6}$ of $SU(3)$. Using the AdS/CFT correspondence, one
identifies this perturbation with a composite operator of 
${\cal N}=3$ superconformal Chern-Simons matter theory with a mass
term for the symmetric and traceless product between 
two $\overline{\bf 3}$'s: $\lambda^{AB} \int d^3 x {\cal O}_{AB}$ where 
$\lambda^{AB}$ is in the ${\bf 6}$ of $SU(3)$. 
Note that the tensor product of these leads to
$
\overline{{\bf 3}}(\tiny\yng(1,1)) \times \overline{{\bf 3}}(\tiny\yng(1,1)) = 
{\bf 3}(\tiny\yng(1)) 
\oplus \overline{{\bf 6}}(\tiny\yng(2,2))$. 
Then one can construct a $\overline{\bf 3}$(that is the
Young tableaux $\tiny\yng(1,1)$)
representation by
using the Clebsch-Gordan coefficient $\Gamma_{AIJ}(A=1, 2, 3)$ 
which transforms
two ${\bf 3}$'s into $\overline{\bf 3}$ of $SU(3)$
(${\bf 3}(\tiny\yng(1)) \times {\bf 3}(\tiny\yng(1)) = 
\overline{\bf 3}(\tiny\yng(1,1)) 
\oplus {\bf 6}(\tiny\yng(2))$) together with matter field
$C^I$: 
$
\Gamma_{AIJ} C^I C^J
$
where
$C^I(I=1, 2, 3)$ are three
complex scalars(${\bf 3}$ under the $SU(3)$) 
transforming as $({\bf N, \overline{N}})$ with gauge
group $SU(N) \times SU(N)$ in ${\cal N}=3$ superconformal Chern-Simons
gauge theory \cite{BFFMZ}.  
The perturbation is given by 
\bea
{\cal O}_{AB} \sim \Tr \Gamma_{AIJ} C^I
C^J \Gamma_{BKL} C^K C^L.
\nonu
\eea
The singlet 
of this operator ${\cal O}_{AB}$ which is $\overline{\bf 6}$ 
of $SU(3)$ corresponds to the
supergravity
field $v(x)$(or $\overline{v}(x)$) and the conformal dimensions are
given by $\Delta_{UV} = \frac{4}{3}$(or $\frac{5}{3}$) and $\Delta_{IR}=4$  
respectively as we computed before.
The other five states with $J=2$ among  $\overline{\bf 6}$ 
of $SU(3)$ are non-diagonal and correspond to deformations of the 
seven-dimensional metric \cite{BFFMZ1}. 
Since the Lichnerowicz operator provides nine-dimensional space,
the remaining three states correspond to the eigenvalue 
$M_{{\bf 200}}=0$. These 
are organized in a triplet$(J=1)$ of $SU(2)_R$ and the massless
scalars
belong to the additional massless vector multiplet.   
So far we considered only 
FR compactification where there are no internal components for the
four-form field strength. In \cite{BFFMZ1}, they further described the
deformation from turning on an internal three-form.   

\section{
Conclusions and outlook }

We have constructed the full eleven-dimensional metric given by 
(\ref{11dmetric}) and obtained  the scalar potential in 
(\ref{potential}) by using the Freund-Rubin ansatz
(\ref{fieldstrength}).  
The holographic supersymmetric(or nonsupersymmetric) RG flow 
from ${\cal N}=(0, 1)$ 
$SU(3) \times SU(2)$-invariant UV fixed
point to ${\cal N}=(3, 0)$ 
$SU(3) \times SU(2)$-invariant IR fixed point was described.
The corresponding operator in three-dimensional 
Chern-Simons matter theories is identified.

For the seven-dimensional Einstein metric, we considered $SO(3)$-bundle over
the base ${\bf CP}^2$ in (\ref{11dmetric}). 
Also the base ${\bf CP}^2$ can be replaced by either ${\bf
S}^4$ or ${\bf CP}^1 \times {\bf CP}^1$. It would be interesting to 
find out whether these metrics provide the new nontrivial  eleven-dimensional 
solutions.  
Are there any new general $AdS_4$ vacua corresponding to any
supersymmetric Chern-Simons matter theories? Are there any new critical
points in the context of $SO(3)$ gauged supergravity(which might be related to 
${\cal N}=3$ Chern-Simons matter theory) or 
$SO(4)$ gauged supergravity?

\vspace{.7cm}
\centerline{\bf Acknowledgments}

This work was supported by grant No.
R01-2006-000-10965-0 from the Basic Research Program of the Korea
Science \& Engineering Foundation.  

%

\end{document}